\documentclass[prb,12pt,tightenlines,amsmath,amssymb]{revtex4}
\usepackage{graphicx}
\usepackage{ulem}

\begin{document}
\title{ \bf  Classification of a supersolid: Trial wavefunctions, Symmetry breakings and Excitation spectra   }
\author{  Yu Chen $^{1}$, Jinwu Ye$^{2,3}$ and Quang Shan Tian $^{1}$ }
\affiliation{ $^{1}$Department of Physics, Peking University, Beijing 100871, China\\
$^{2}$  Department of Physics and Astronomy,
Mississippi State University, P. O. Box 5167, Mississippi State, MS, 39762   \\
$^{3}$ Department of Physics, Capital Normal University, Beijing, 100048 China  }
\date{\today}

\begin{abstract}
   A state of matter is characterized by its symmetry breaking and elementary excitations.
   A supersolid is a state which breaks both translational symmetry and internal $ U(1) $ symmetry.
   Here, we review some past and recent works in phenomenological Ginsburg-Landau theories, ground state trial wavefunctions  and
   microscopic numerical calculations. We also write down a new effective supersolid Hamiltonian on a lattice.
   The eigenstates of the Hamiltonian contains both the ground state wavefunction and all the excited states ( supersolidon ) wavefunctions.
   We contrast various kinds of supersolids in  both continuous systems and on lattices, both condensed matter and cold atom  systems.
   We provide additional new insights in studying their order parameters, symmetry breaking patterns, the excitation spectra and detection methods.

\end{abstract}

{\sl  Contribution to the supersolid JLTP special issue   SS2012 }

\maketitle


\section{ Introduction }

     A solid can not flow. It breaks a continuous translational
     symmetry into a discrete lattice translational symmetry.
     There are low energy lattice phonon excitations in the solid.
     While a superfluid can flow even through narrowest channels
     without any resistance. It breaks a global $ U(1) $  symmetry and has the off-diagonal long range order (ODLRO). There are low energy superfluid phonon excitations
     in the superfluid. A supersolid is a state which breaks both
     the continuous translational symmetry and the global $ U(1) $
     symmetry, therefore has both the crystalline  order and the
     ODLRO.  The possibility of a supersolid phase in $^{4}He$ was
     theoretically speculated in 1970 \cite{and,ches}.
     If so, under the slow rotation of a container, the superfluid component can not rotate, therefore reduces the rotational moment of inertial.
     This reduction is called Non-Classical Rotational Inertia (NCRI) \cite{leg}. Over the last 40
     years, a number of experiments have been designed to search for
     the supersolid state. Most notably, by
     using torsional oscillator measurements, the group led by Chan observed
     a marked $ 1 \sim 2 \% $ NCRI of solid $^{4}He$
     at $ \sim 0.2 K $ in bulk $^{4}He$ \cite{chan}.
     The authors suggested that the supersolid state of $^{4}He$ maybe responsible
     for the NCRI. The experiments rekindled extensive
     both experimental \cite{annealing,massflow,massflow1,melt,acou,heat,xray} and theoretical \cite{micro,micro2,ander,ander2,dor1,dor2,son,qglprl,epl,qgllong,noz,miss1,disorder,entropy} interests.
     So far, there is still a controversy if a supersolid phase indeed exist in $ ^{4} He $ system and is responsible for
     the NCRI observed in the Chan's experiments. For example, the NCRI experiment in annealed samples \cite{annealing}
     the mass flow experiments \cite{massflow,massflow1} and the first principle microscopic calculations \cite{micro}
     indicate that the superfluid effects observed so far in solid He-4 are disordered-induced.

     In this manuscript, instead of trying to resolve this controversy,
     we will discuss some universal properties of a supersolid
     such as its ground state wavefunctions, symmetry breakings, elementary excitation spectra and their detections in various possible experimental systems.
     No matter if supersolid indeed exists in the  $ ^{4} He $ system or not, it is a new state of matter having its own characteristic  behaviors
     not shared by any other states of matter.  Various kinds of supersolids may also be realized in other various bosonic or fermionic, continuous
     or optical lattice systems.

     The rest of the paper is organized as following. In Sect.II, we will discuss the supersolids in continuous 3d and 2d systems.
     In Sec.II-A, we review the elementary excitation spectrum ( called supersolidon )
     in a possible 3d $^{4}He $ supersolid with Van der Walls interaction. Sec.II-B is new where we
     make connections to vacancy supersolid wave functions and derive an effective supersolid Hamiltonian on lattice scales.
     The supersolidon spectrum discussed in previous works can be extended to the whole Brillouin Zone (BZ).
     In Sec.II-C, we discuss a possible exciton supersolid with dipole-dipole interaction in
     a 2d electron-hole bilayer system in some intermediate distances. Sect. III is dedicated to supersolids in lattice systems.
     We discuss some analogy and also important differences between the continuous supersolids and lattice supersolids.
     We stress that there is a new kind of supersolid in lattice systems: the valence bond supersolid.
     In Sect.IV, we study the superfluid density waves.
     Namely, an inhomogeneous superfluid in a continuous system in IV-A and
     a $ Z_2 $  superfluid density wave inside an optical cavity.
     In the final Sec.V, we contrast different kinds of supersolids addressed in this paper and also summarize our main results.
     The possible important effects of disorders will not be discussed in the manuscript and are referred to the original
     literatures in \cite{annealing,massflow,massflow1,micro,entropy,disorder}


\section{ Supersolids in continuous 3 and 2 dimensional systems }

   Classical non-equilibrium hydrodynamics inside a SS was investigated for a long
   time \cite{and,dor1,dor2,son}. The classical hydrodynamics will break down at very low temperature where quantum fluctuations dominate.
   The quantum phenomenological Ginsburg-Landau theories \cite{qglprl,epl,qgllong} in both 3d and 2d were written down.
 They can be applied to  study possible supersolids in 3 dimensional $ ^{4}He $ system and possible exciton supersolids in 2 dimensional
 electron-hole bilayer systems. The main purposes for the GL theory are
 (1) It can be used to analyze the stability conditions of a supersolid. As shown in \cite{qgllong},
     depending on the parameters of the GL action, the supersolid can be either absent or present in the pressure $ P $ versus
     temperature $ T $ phase diagram. The parameters should be determined by microscopic details of the atom-atom interactions.
 (2) Assuming the supersolid is present in the $ P-T $ phase diagram, at the mean field level,
     the GL theory can be used to determine the lattice structure of a supersolid,
 (3) When considering fluctuations above the mean field solutions, one may study transitions among different phases using the renormalization group analysis.
 (4) Well inside a given phase, especially inside a supersolid, one can study the elementary excitations inside such a given phase.
     The phonon spectrum in a solid or the superfluid phonon ( or Goldstone mode )
     in a superfluid has all been detected by inelastic neutron scattering experiments.
     So detecting these "supersolidons " by possible inelastic neutron scattering experiments or acoustic attenuation experiments
     could be smoking gun experiments to confirm a supersolid in any continuous systems.
 (5) One can study any interacting Bose systems from both path integral quantization and  canonical  quantization approach.
     Both approaches are complementary to each other.
     When combining the effective path integral action inside a supersolid with the known trial vacancy supersolid wavefunctions,
     one can write down an effective Hamiltonian on a lattice to gain additional insights to the physical picture of a supersolid.
     The eigenstates of the Hamiltonian contains both the
     ground state wavefunction and all the excited states ( supersolidon ) wavefunctions.
     The supersolidon spectrum can also be worked out in the whole Brillouin Zone (BZ) including close to the  BZ boundary.

   Here, we will review the nature of low energy excitations in the SS, namely, focus on the (4) in the above.
   We refer (1)-(3) to the original papers \cite{qglprl,epl,qgllong,exss}. The part (5) is new and will be discussed in Sect.II-B.

\subsection{  Possible supersolid in 3 dimensional $ ^{4} He $ with Van der Waals force.   }

     Well inside the SS, the translational symmetry is already broken,
  so we can parameterize the density deviation order parameter
  $ \delta n (\vec{x}, \tau ) = n( \vec{x}, \tau ) - n_{0}  $ and the SF complex
  order parameter $  \psi ( \vec{x}, \tau ) $  as:
\begin{eqnarray}
 \delta n( \vec{x}, \tau ) & = &  \sum^{\prime}_{\vec{G}} n_{\vec{G}} e^{i \vec{G} \cdot
     ( \vec{x} + \vec{u}( \vec{x}, \tau ) ) }   \nonumber  \\
 \psi ( \vec{x}, \tau ) & = &  \psi_{0} ( \vec{x}, \tau ) [  1  \pm \frac{1}{P} \sum^{\prime}_{\vec{G}}
      e^{i \vec{G} \cdot
     ( \vec{x} + \vec{u}( \vec{x}, \tau ) )} ]
\label{orderhe}
\end{eqnarray}
     where the $ \psi_{0}( \vec{x}, \tau )= | \psi_{0}( \vec{x}, \tau ) | e^{ i \theta(  \vec{x}, \tau ) }  $
     is the SF order parameter, $ \vec{u}( \vec{x}, \tau )  $ are the 3 lattice phonon modes,
     the $ \pm $ means vacancy-type or interstitial-type supersolids respectively,
     $ n^{*}_{\vec{G}}= n_{-\vec{G}} $ the " $\prime $ " means the sum over the shortest non-zero
     reciprocal lattice vector  $ \vec{G} $ and $ P $ is the number of them.

   The long wavelength effective action describing the low energy  modes inside the SS phase was derived in \cite{epl,qgllong}:
\begin{equation}
   {\cal L}_{SS}   =   \frac{1}{2} [ \kappa ( \partial_{\tau} \theta )^{2} +
    \rho^{s}_{\alpha \beta} \partial_{\alpha} \theta \partial_{\beta}
    \theta]     +  \frac{1}{2}[ \rho_{n}  ( \partial_{\tau} u_{\alpha} )^{2} +
    \lambda_{\alpha \beta \gamma \delta} u_{\alpha \beta
    } u_{\gamma \delta } ] + a_{\alpha \beta }  u_{\alpha \beta } i  \partial_{\tau} \theta
\label{ss}
\end{equation}
    where  $ \kappa $ is the SF compressibility  and $ \rho^{s}_{\alpha \beta} $
    is the SF stiffness which has the same symmetry as $  a_{\alpha \beta }  $. The $ \rho_n $ is the normal density,
    the $ u_{\alpha \beta }= \frac{1}{2}( \partial_{\alpha}
    u_{\beta} + \partial_{\beta} u_{\alpha} ) $ is the strain
    tensor, the $ \lambda_{\alpha \beta \gamma \delta} $ is the elastic constant tensor.
    Obviously, the last term is the crucial {\sl Berry phase }
    coupling term which couples the lattice phonon modes to the
    SF mode. The factor of $ i $ is important in this coupling.
    By integration by parts, this term can also be written as $ a_{\alpha
    \beta} ( \partial_{\tau} u_{\beta} \partial_{\alpha} \theta
    + \partial_{\tau} u_{\alpha} \partial_{\beta} \theta ) $ which
    has the clear physical meaning of the coupling between the SF
    velocity $ \partial_{\alpha} \theta $ and the velocity of
    the lattice vibration $ \partial_{\tau} u_{\beta} $.
    It is this term which makes
    the low energy modes in
    the SS to have its own characteristics which could be detected by
    experiments. The invariance under the Galilean transformation \cite{dor2} dictates that $ a_{\alpha,\beta}=\rho_n \delta_{\alpha,\beta}- \rho^{s}_{\alpha,\beta} $. Here, we only review the isotropic solid case,  the $ hcp $ lattice and the effects of the topological
    vortex loop excitations were discussed in the original papers \cite{epl,qgllong}.

    A truly isotropic solid can only be realized in a highly poly-crystalline
    sample. Usual samples are not completely isotropic. However, we
    expect the simple physics brought about in an isotropic solid
    may also apply qualitatively to other samples which is very
    poly-crystalline.   For an isotropic solid, $ \lambda_{\alpha \beta \gamma \delta}=
    \lambda \delta_{\alpha \beta} \delta_{\gamma \delta}
    + \mu ( \delta_{\alpha \gamma } \delta_{\beta \delta}+ \delta_{\alpha \delta} \delta_{\beta \gamma
    } ) $ where $ \lambda $ and $ \mu $ are Lame coefficients, $
    \rho^{s}_{\alpha,\beta}= \rho^{s} \delta_{\alpha,\beta}, a_{\alpha,\beta}= a
    \delta_{\alpha,\beta} $ where  $ a= \rho_n- \rho_s $. In $ ( \vec{q}, \omega_{n} ) $ space, Eqn.\ref{ss} becomes:
\begin{eqnarray}
    {\cal L}_{ISS} & = &  \frac{1}{2}[ \rho_{n} \omega^{2}_{n} + ( \lambda+2 \mu
    ) q^{2} ] |u_{l}( \vec{q},\omega_{n} ) |^{2}   +  \frac{1}{2} [ \kappa \omega^{2}_{n} + \rho_{s} q^{2} ] |\theta ( \vec{q},\omega_{n} ) |^{2}
                    \nonumber  \\
    &  + &  a q \omega_{n} u_{l}( -\vec{q}, - \omega_{n} ) \theta ( \vec{q},\omega_{n} )
    +   \frac{1}{2}[ \rho_{n} \omega^{2}_{n} + \mu q^{2} ] |u_{t}( \vec{q},\omega_{n} ) |^{2}
\label{is}
\end{eqnarray}
     where $ u_{l}( \vec{q},\omega_{n} )= i q_{i} u_{i}(
     \vec{q},\omega_{n} )/q $ is the longitudinal component,
     $ u_{t}( \vec{q},\omega_{n} )= i \epsilon_{ij} q_{i} u_{j}( \vec{q},\omega_{n}
     )/q $ are transverse components of the
     displacement field. Note that Eqn.\ref{is}
     shows that only longitudinal component couples to the
     superfluid  $ \theta $ mode, while the two transverse components
     are unaffected by the superfluid mode. This is expected, because
     the superfluid mode is a longitudinal density mode itself which
     does  not couple to the transverse modes.

\begin{figure}
\includegraphics[width=6cm]{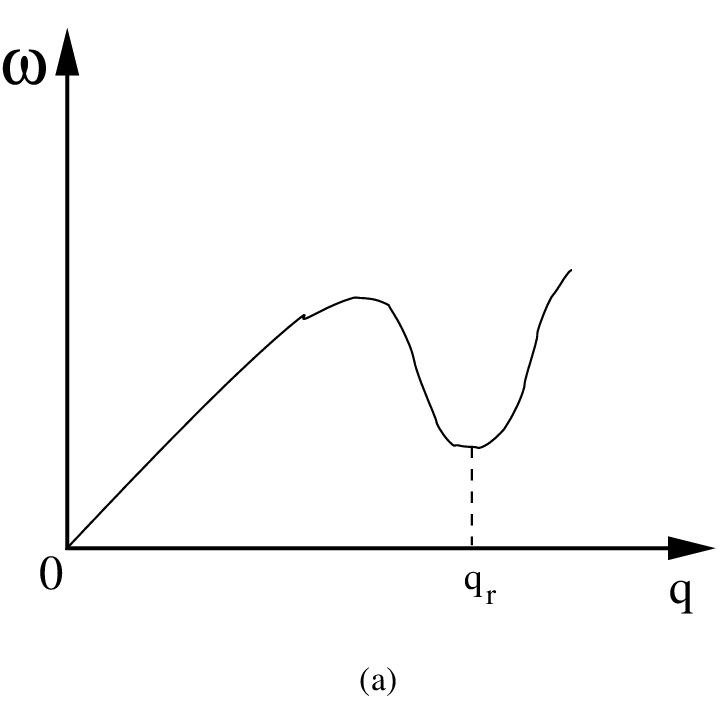}
\hspace{0.5cm}
\includegraphics[width=8cm]{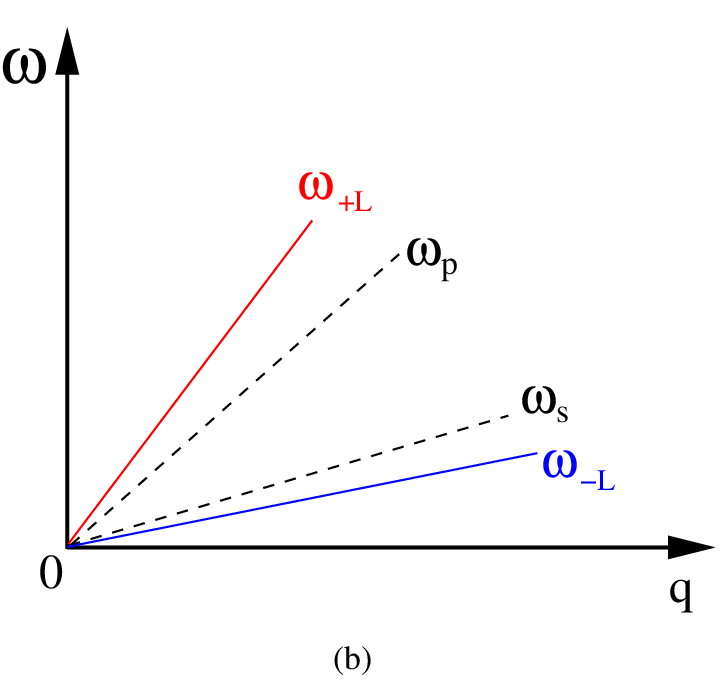}
\hspace{0.5cm}
\includegraphics[width=6cm]{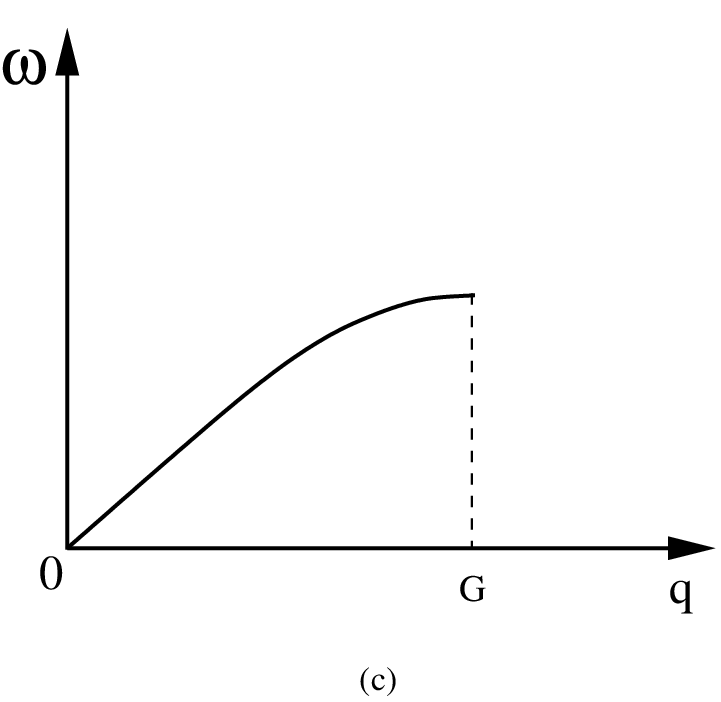}
\caption{ (Color figure online) The elementary low energy excitations inside a (a) superfluid, the $ q_r \sim G $ is the roton minimum position.
(b) supersolid (c) solid, the $ G $ is the shortest reciprocal lattice vector in the Brillouin Zone (BZ).
For simplicity, only excitation spectra in a simplified 3d isotropic solid and 2d triangular lattice
were shown. In (b) The coupling between the phonon mode $ \omega_{p} = v_{p}q $ ( the
upper dashed line ) and the superfluid mode $ \omega_{s}=v_{s}q $ (
the lower dashed line ) leads to the two new "supersolidon " modes $
\omega_{\pm}=v_{\pm} q $ when $ q \ll G  $ ( solid lines ) in the SS. Their corresponding spectral weights are  listed in Eqn.\ref{uu0} and \ref{tt}.
These two new supersolid modes and the spectral weights should be detected by in-elastic neutron
scattertings \cite{mermin}. The dispersion form of the supersolidon near BZ $ G $ can be worked out from the effective Hamiltonian at
the lattice scale Eqn.\ref{sslattice}.  } \label{fig6}
\end{figure}

     From Eqn.\ref{is}, we can identify the
     longitudinal-longitudinal phonon correlation function:
\begin{equation}
  \langle u_{l} u_{l} \rangle= \frac{ \kappa \omega^{2}_{n} + \rho_{s} q^{2} }{
  ( \kappa \omega^{2}_{n} + \rho_{s} q^{2} ) (  \rho_{n} \omega^{2}_{n} + ( \lambda+2 \mu
    ) q^{2} ) + a^{2} q^{2} \omega^{2}_{n} }
\end{equation}
     The $ \langle \theta \theta \rangle $ and $ \langle u_{l} \theta \rangle $ correlation
     functions can be similarly written down. By doing the
     analytical continuation $ i \omega_{n} \rightarrow \omega + i
     \delta $, we can identify the two poles of  all the correlation
     functions at $ \omega^{2}_{\pm}= v^{2}_{\pm} q^{2},~~~ q \ll G $ where the
     two velocities $ v_{\pm} $ is given by:
\begin{equation}
  v^{2}_{\pm}  = [ \kappa( \lambda+2 \mu ) + \rho_{s} \rho_{n} +
   a^{2} \pm  \sqrt{ ( \kappa( \lambda+2 \mu ) + \rho_{s} \rho_{n} +
   a^{2} )^{2}- 4 \kappa \rho_{s} \rho_{n} ( \lambda+2 \mu ) } ]/ 2 \kappa \rho_{n}
\end{equation}
    If setting $ a =0 $, then $ v^{2}_{\pm} $ reduces to the longitudinal phonon
    velocity $ v^{2}_{lp}=  ( \lambda + 2 \mu )/ \rho_{n} $ and
    the superfluid velocity $ v^{2}_{s} = \rho_{s}/\kappa $ respectively.
    Of course, the transverse phonon velocity $ v^{2}_{tp}= \mu /
    \rho_{n} $ is untouched. For notation simplicity, in the
    following, we just use $ v_{p} $ for $ v_{lp} $.
    Inside the SS, due to the very small superfluid density $
    \rho_{s} $, it is expected that $ v_{p} > v_{s} $.
    In fact, in isotropic solid $ He^{4} $, it was measured that $ v_{lp} \sim 450-500
    m/s, v_{t} \sim 230 \sim 320 m/s $ and $ v_{s} \sim 366 m/s $ near the melting curve \cite{melt}.
    It is easy to show that $ v_{+} > v_{p}> v_{s} > v_{-} $ and
    $ v^{2}_{+} + v^{2}_{-}  >  v^{2}_{p} + v^{2}_{s} $, but  $ v_{+} v_{-} = v_{p} v_{s} $,
    so $ v_{+} + v_{-} > v_{p} + v_{s} $ ( see
    Fig.1 ). Note that because the Galilean invariance dictates $ a= \rho_n- \rho_s $,
    for $ \rho_s \ll \rho_n $, one can see $ \rho_{s} \rho_{n} +   a^{2} \gg \rho_{s} \rho_{n} $, so
    $ v_{+} $ ( $ v_{-} $ ) are considerably above ( below ) $ v_{p} $ ( $ v_{s} $ ), so the two supersolidons,
    especially the softening of the lower branch, may be
    easily distinguished by possible neutron scattering experiments.

    By doing the  analytical continuation $ i \omega_{n} \rightarrow \omega + i
     \delta $, we can take the imaginary part and find:
\begin{eqnarray}
     Im  \langle u_{l} u_{l} \rangle_{i \omega_{n} \rightarrow \omega + i
     \delta}  & = & \frac{ v^{2}_{s}- v^{2}_{+} }{  v^{2}_{+}- v^{2}_{-} } \frac{ \pi}{ 2 \rho_{n} v_{+} } \frac{1}{q}
     [ \delta( \omega- v_{+}q ) -  \delta( \omega+ v_{+}q ) ]   \nonumber  \\
     & - & \frac{ v^{2}_{s}- v^{2}_{-} }{  v^{2}_{+}- v^{2}_{-} } \frac{ \pi}{ 2 \rho_{n} v_{-} } \frac{1}{q}
     [ \delta( \omega- v_{-}q ) -  \delta( \omega+ v_{-}q ) ]
\label{uu0}
\end{eqnarray}
    It is easy to see that the second term can be achieved from the first term just by $ v_{+} \leftrightarrow v_{-} $.
    Setting $ a=0 $, then $ v_{+} = v_{p} $, $ v_{-}=v_{s} $, the second term just vanishes, the first term recovers
    the excitation spectrum of the lattice phonons. When $ a \neq 0 $, then Eqn.\ref{uu0} becomes a mixing of the lattice phonons and superfluid
    phonons, the first and second term give the excitation energies and the two corresponding spectral weights.

    Very similarly, we can find
\begin{eqnarray}
     Im  \langle \theta \theta \rangle_{i \omega_{n} \rightarrow \omega + i
     \delta}  & = & \frac{ v^{2}_{p}- v^{2}_{+} }{  v^{2}_{+}- v^{2}_{-} } \frac{ \pi}{ 2 \kappa v_{+} } \frac{1}{q}
     [ \delta( \omega- v_{+}q ) -  \delta( \omega+ v_{+}q ) ]   \nonumber  \\
     & - & \frac{ v^{2}_{p}- v^{2}_{-} }{  v^{2}_{+}- v^{2}_{-} } \frac{ \pi}{ 2 \kappa v_{-} } \frac{1}{q}
     [ \delta( \omega- v_{-}q ) -  \delta( \omega+ v_{-}q ) ]
\label{tt}
\end{eqnarray}
    It is easy to see that the second term can be achieved from the first term just by $ v_{+} \leftrightarrow v_{-} $.
    Setting $ a=0 $, then $ v_{+} = v_{p} $, $ v_{-}=v_{s} $, the first term just vanishes, the second term recovers
    the excitation spectrum of the superfluid phonons. When $ a \neq 0 $, then Eqn.\ref{tt} become a mixing of the lattice phonons and superfluid
    phonons, the first and second term give give the excitation energies and the two corresponding spectral weights.
    So detecting these "supersolidons " in Fig.1b by possible inelastic neutron scattering experiments or acoustic attenuation experiments
    could be smoking gun experiments to confirm a supersolid in any continuous systems. The experimental implications of the
    supersolidons on X-ray scatterings, density-density correlation functions, specific heats wee discussed in \cite{epl,qgllong}.

    However, it is well known that the GL theory is a phenomenological theory.
    In fact, depending on the parameter regimes in the GL theory in \cite{qgllong}, the author discussed both
    the non-existence and existence of supersolid, also the vacancy-type and interstitial-type the supersolid  if it exists.
    Into which parameter regime will the $^{4} He $ fall can only be studied by various numerical calculations \cite{micro,micro2}.
    But so far, it seems that most of the numerical simulations with the $^{4} He $ atoms interacting each other with Van der Waals forces
    favor a commensurate solid instead of either vacancy-type or interstitial type supersolid.
    Even so, the supersolid could be a meat-stable phase with sufficiently long life time and lead to
    some experimental observable signatures within some time scale.

\subsection{ Connections with the trial Wavefunctions of a supersolid  and an effective Hamiltonian }

 In this subsection, we contrast wave functions of a conventional solid, a solid with local tunneling process and a supersolid.
 We also contrast their corresponding  effective Hamiltonian.

{\sl 1. Trial wavefunction for a supersolid and the effective Hamiltonian }

    A well known trial wavefunction \cite{ches,micro,micro2} for a supersolid in the second quantization form was written in the appendix A of Ref.\cite{qgllong}:
\begin{equation}
    |SS, \phi\rangle = \prod^{N}_{i=1} ( \cos \frac{\theta}{2} + \sin \frac{\theta}{2}
    e^{i \phi}  b^{\dagger}_{i} ) |0\rangle
\label{ssp}
\end{equation}
    where $ \theta \neq \pi $. The average boson number is $ M=N \sin^{2} \frac{\theta}{2} < N  $.
    The vacancy number is $ N_b=N \cos^{2} \frac{\theta}{2} >0 $.

     A supersolid state $ |SS, M \rangle $ with $ M < N  $ bosons is given by:
\begin{equation}
    |SS, M \rangle  =  \int^{2 \pi}_{0} \frac{ d \phi}{2
     \pi} e^{-i M \phi} |SS, \phi\rangle
     =   ( \cos \frac{\theta}{2} )^{N} C^{M}_{N} ( \tan \frac{\theta}{2} )^{M}
    \frac{1}{ M !} \sum_{ i_1,\cdots,i_{M} }  b^{\dagger}_{i_1} \cdots  b^{\dagger}_{i_{M}}  |0\rangle
\label{ssm}
\end{equation}
     where the total boson number $ M $ and the global phase $ \phi $ are two Hermitian conjugate
     variables satisfying the commutation relation: $ [ \delta M, \phi]=i \hbar  $.
     It leads to the uncertainty relation $ \Delta M \Delta \phi \geq 1 $.

     Following the procedures in \cite{wavefunction} to derive the first quantization form of a wavefunction from its second quantized form
     in the exciton superfluid in bilayer quantum Hall systems, one can find the first quantization  form of the supersolid wavefunction:
\begin{equation}
     |SS, M \rangle =  C \sum_{P} {\cal S} [\prod^{M<N}_{i=1} \phi( \vec{r}_i-\vec{R}_{P_i})]
\end{equation}
     where the $ P $ is the sum over all the $ C^{M}_{N} $ possible ways of selecting $ M < N $ sites from the $ N $ sites.
     The $ {\cal S } $ is the symmetrization acting on the boson coordinates $ \vec{r}_{i} $. The $ C $ is the normalization constant.

      In order to consider the mutual interaction between atoms in neighboring sites, it is necessary to incorporate the Jastrow factor
      \cite{ches} into the above wave function:
\begin{equation}
      |SS, M \rangle_{J} =  C \sum_{P} {\cal S} [\prod^{M< N}_{i=1} \phi( \vec{r}_i-\vec{R}_{P_i} ) \prod_{i < j } J( r_{ij} ) ]
\label{ssj}
\end{equation}
       where the $ \prod_{i < j } J( r_{ij} ) = e^{- \sum_{i < j} u( r_{ij} ) } $ is the Jastrow factor, the $ u( r_{ij} ) $
       is the Lennard-Jones 6-12 potential with a hard core repulsion.

       When comparing the effective action Eqn.\ref{ss} with the ground state wavefunction Eqn.\ref{ssp}, one can see that
       the angle $ \theta $ corresponds to the superfluid density fluctuation,
       while the angle $ \phi $ is the most important phase fluctuations in Eqn.\ref{ss}. The phonon modes in \ref{ss} corresponds to
       $ \vec{r}_{i} \rightarrow \vec{R}_{i} + \vec{u} ( \vec{R}_{i} ) $ in Eqn.\ref{ssj}. In fact, as argued in the following,
       the trial wavefunction Eqn.\ref{ssj} can be taken as the exact ground state of the effective Hamiltonian Eqn.\ref{sslattice}.

       Motivated by the one to one correspondence between the effective action Eqn.\ref{ss} and the ground state wavefunctions Eqn.\ref{ssj},
       one can write the corresponding effective Hamiltonian on a lattice scale:
\begin{eqnarray}
       {\cal H}_{SS}= \sum_{\vec{R} \in M < N } \frac{ P^{2} (\vec{R} ) }{ 2 M } + \frac{1}{2} \sum_{\alpha, \beta, \vec{R},\vec{R}^{\prime} }
       u_{\alpha} (\vec{R} ) D_{\alpha \beta} ( \vec{R} - \vec{R}^{\prime} ) u_{\alpha} (\vec{R}^{\prime} )  \nonumber  \\
       + \frac{1}{2} \delta \rho (-\vec{q} ) V(\vec{q} ) \delta \rho (\vec{q} ) +  \frac{1}{2} \rho^{s}_{\alpha \beta} \partial_{\alpha} \theta \partial_{\beta}
       \theta + a_{\alpha, \beta } ( \Delta_{\alpha} u_{\beta} + \Delta_{\beta} u_{\alpha } ) \delta \rho
\label{sslattice}
\end{eqnarray}
      where the sum over $ \vec{R} \in M < N $ is only over the $ M < N $ sites where the atoms occupy,
      the vacancy number $ b^{\dagger} b = N-M $ flow though the whole lattice and condense into the superfluid state.
      The SF compressibility $ \kappa^{-1} = lim_{ q \rightarrow 0 } V(\vec{q} ) $ where the form $ V(q) = a -bq^{2}+ c q^{4}, a,b,c > 0 $ \cite{qgllong}
      is needed to lead to the superfluid mode in Fig 1a in the whole BZ. The $ u_{\alpha}, P_{\beta} $ and the $ \theta, \delta \rho $ are the two sets of conjugate variables
      satisfying $ [ u_{\alpha} ( \vec{R} ),  P_{\beta} ( \vec{R}^{\prime} ) ]= i \hbar \delta_{\alpha, \beta} \delta_{ \vec{R}, \vec{R}^{\prime} } $ and
      $ [ \theta, \delta \rho ]= i \hbar $. When the vacancies $ b $ are moving through the whole lattice \cite{hightc}, the superfluid density fluctuation couples  to the lattice vibration through the last term where $ \Delta_{\alpha} u_{\beta} = u_{\beta} ( \vec{R} + \hat{e}_{\alpha} )- u_{\beta} ( \vec{R} ) $ is the lattice difference.

      Obviously, the long wavelength limit of Eqn. \ref{sslattice} leads to Eqn.\ref{ss}. The effective Hamiltonian  also hold on lattice scales.
      When neglecting the vortex excitations, the Eqn.\ref{sslattice} is quadratic,
      the standard Bogliubov transformation can be used to extend  the supersolidon dispersion
      relation in Fig.1b to the whole BZ zone including near to the BZ boundary $ G $ in the Fig 1c.
      The vortex excitations inside a supersolid were discussed in \cite{qgllong}.

      For comparisons, in the following, we also write down the wavefunction and the corresponding effective Hamiltonian of a
      conventional solid and a quantum solid with local quantum tunneling and exchange process.

{\sl 2.  Wavefunction for a conventional solid and the effective Hamiltonian }

       For a conventional solid, one atom is attached to a given lattice site. Because all the atoms can be treated as distinguishable as
       labeled by its attached site $ i $, so the symmetrization operator $ {\cal S} $ in Eqn.\ref{ssj} is not necessary.
\begin{equation}
      |Solid \rangle = \prod^{M}_{i=1} \phi( \vec{r}_i-\vec{R}_{i} ) \prod_{i < j } J( r_{ij} )
\label{solid}
\end{equation}
      Obviously, its effective Hamiltonian is just the first line in the Eqn.\ref{sslattice}.
\begin{equation}
       {\cal H}_{Solid}= \sum_{\vec{R} \in M < N } \frac{ P^{2} (\vec{R} ) }{ 2 M } + \frac{1}{2} \sum_{\alpha, \beta, \vec{R},\vec{R}^{\prime} }
       u_{\alpha} (\vec{R} ) D_{\alpha \beta} ( \vec{R} - \vec{R}^{\prime} ) u_{\alpha} (\vec{R}^{\prime} )
\label{solidh}
\end{equation}
       whose phonon spectrum has been discussed in textbooks

{\sl 3.  Wavefunction for a commensurate solid with local tunnelings and its effective Hamiltonian }

       In a commensurate solid, the number of atoms is equal to the number of sites $ M=N $. There exist still local quantum tunneling  and exchange processes.
       But these local quantum tunneling and exchange process will not lead to a global phase coherence.
       Due to these local tunneling processes, the atoms still need to be treated as in-distinguishable identical particles,
       so the the symmetrization operator $ {\cal S} $ in Eqn.\ref{ssj} is still necessary, the wavefunction can be written as
\begin{equation}
      |Solid \rangle_{le} =  {\cal S} [\prod^{M= N}_{i=1} \phi( \vec{r}_i-\vec{R}_{P_i} ) \prod_{i < j } J( r_{ij} ) ]
\label{solidle}
\end{equation}
       where the $ le $ means local exchange processes. The corresponding Hamiltonian is
\begin{eqnarray}
       {\cal H}_{solid+le}= \sum_{\vec{R} \in M < N } \frac{ P^{2} (\vec{R} ) }{ 2 M } + \frac{1}{2} \sum_{\alpha, \beta, \vec{R},\vec{R}^{\prime} }
       u_{\alpha} (\vec{R} ) D_{\alpha \beta} ( \vec{R} - \vec{R}^{\prime} ) u_{\alpha} (\vec{R}^{\prime} )  \nonumber  \\
       - \frac{ \hbar^{2} }{ 2 m } \psi^{\dagger} \nabla^{2} \psi + \mu  \psi^{\dagger} \psi + u  ( \psi^{\dagger} \psi )^{2} + a_{\alpha, \beta } ( \Delta_{\alpha} u_{\beta} + \Delta_{\beta} u_{\alpha } ) \psi^{\dagger} \psi
\label{solidleh}
\end{eqnarray}
      where  the $ u_{\alpha}, P_{\beta} $ and the $ \psi, \psi^{\dagger} $ are the two sets of  conjugate variables
      satisfying $ [ u_{\alpha} ( \vec{R} ),  P_{\beta} ( \vec{R}^{\prime} ) ]= i \hbar \delta_{\alpha, \beta} \delta_{ \vec{R}, \vec{R}^{\prime} } $ and
      $ [ \psi ( \vec{x} ) , \psi^{\dagger}( \vec{x}^{\prime} ) ]= i \hbar \delta( \vec{x}-\vec{x}^{\prime} ) $.
      The local tunneling mode couples  to the lattice vibration through the last term. Due to the positive mass term $ \mu > 0 $, the local mode has a gap,
      so integrating out the gapped mode $ \psi $ will lead to the same effective Hamiltonian Eqn.\ref{solidh} as the conventional solid.
      So the lattice phonon mode will not
      be affected by the gapped bulk normal fluid mode.

\subsection{  Possible 2 dimensional Exciton supersolid in electron-hole bilayer with dipole-dipole interaction.   }

    In this subsection, we will discuss another bosonic system with  much longer range dipole-dipole interactions:
    excitons in electron-hole semi-conductor bilayer (EHBL) systems \cite{exss}.
    We will argue that due to the special form of dipole-dipole interaction, this system may have a better chance to realize a supersolid in some parameter regime.
    There are also some numerical evidences to support such a claim \cite{eplss}. Indeed, as shown in \cite{qgllong} and briefly mentioned at the beginning of
    Sect.I, depending on the parameters of the GL action, the supersolid can be either absent or present in the pressure $ P $ versus
    temperature $ T $ phase diagram. The parameters, in turn, are determined by microscopic details of the atom-atom interactions.

    In the last decade, degenerate exciton systems have been produced by different experimental groups
    with different methods  in quasi-two-dimensional semiconductor $ GaAs/AlGaAs $ coupled quantum wells structure \cite{exss}.
    There are two important dimensionless parameters in the EHBL. (1) One is
the dimensionless distance $ \gamma= d/a_{ex} $   between the two layers.
The   $ a_{ex}=\hbar ^{2}\epsilon /e^{2}m_{r} = \epsilon \frac{m_{0}}{m_{r}} a_{B} \sim 100
a_{B} \sim 50 \AA $ is the size of an exciton, the $ m_0 $ is the electron bare mass and  $ 1/m_{r}=1/m_{e} + 1/m_{h} $
is the reduced mass of the excitons and $ a_{B}=\hbar ^{2} /e^{2}m_{0} \sim 0.53
\AA $ is the bare Bohr radius. The binding energy of an exciton is $ E_{b}= -e^{2}/2a_{ex}\epsilon = -
\frac{m_{r}}{m_{0}} \frac{1}{ \epsilon^{2}}  \frac{ e^{2}}{ 2 a_{B}} \sim -10 meV $.
(2) Another is $ r_s $. The $ r_{s} a_{ex} $, defined by $ \pi (r_{s} a_{ex})^{2} n =1 $,  is the typical interparticle
distance in a single layer. The $ r_{s} $ is the ratio of the  the potential energy over kinetic energy in a
single layer. It is easy to see that the ratio of intralayer Coulomb $ V_{11} $ over the interlayer Coulomb $ V_{12} $
    interaction is given by  $  \alpha =  V_{11}/V_{12} = d/ r_{s} a_{ex} $.
    So when the interlayer Coulomb interaction dominates  $ \alpha < 1
    $, the EHBL is expected to exhibit the superfluid of excitons.
If the density of excitons is sufficiently low (  large $ r_{s} $ ),
then the system is in a weakly coupled Wigner solid  state at very
     large distance and become an BEC excitonic superfluid ( ESF ) at short
     distance.  In the following, we argue \cite{exss} that there could be an exciton supersolid (ESS) phase intervening
     between the ESF and Wigner solid phase, as the system evolves
     from the BEC ESF to the weakly coupled Wigner solid when the distance increases.
     The argument relies heavily on the dipole-dipole interaction between the excitons.

    If we assume an exciton is already formed at relatively short interlayer distance,
    its kinetic energy $ K \sim \frac{ \hbar^{2} }{ M (r_{s}
    a_{ex} )^{2} } $ where $ M = m_{e}+ m_{h} $ is the total mass of an exciton. Due to the dipole-dipole interactions between the excitons,
    its potential energy $ P \sim \frac{ e^{2}  d^{2} }{ \epsilon ( r_{s} a_{ex} )^{3} } $. When $ K < P $,
    namely, $ \sqrt{ m_{r}/M }\sqrt{r_{s}}  <  d/a_{ex}  $,
    the EHBL could favor a excitonic ( or dipolar) normal solid (ENS) state.
    As argued above, when $ d/a_{ex} < r_{s} $, the EHBL is in a ESF state.  So
    in the intermediate distance $ \sqrt{r_{s}}/2 < d/a_{ex} < r_{s} $ where we used $ M/m_{r} \sim 5 $,
    the system may favor a excitonic ( or dipolar ) supersolid (ESS) state.
    When $ d/a_{ex} > r_{s}  $, it will become the excitonic normal solid (ENS)
    due to the long range dipole-dipole $ 1/r^{3} $ repulsive interactions.
    For the present experimental density regime \cite{exss} $ n \sim 10^{10} cm^{-2} $, $ r_{s} \sim 20 $, so the excitons are
    tightly bound pairs in real space.  In this $ r_{s} \sim 20 \gg 1 $ limit, there is a broad distance regime $ 2 < d/a_{ex} < 20
    $ where the system could be in the ESS state. As the distance increases to the critical
     distance $ d > d_{c1} $, the roton minimum in the Fig.1a may drive the instability of the ESF to a formation of a solid.
     Because the lattice constant $ r_{s} a_{ex} $ of the resulting solid is
     completely fixed by the parameter $ r_{s} $ which is independent of the distance, so the resulting solid
     is likely to have vacancies with density  $ n_{v}(0) $ even at $ T=0 $.
     By contrast, in solid Helium 4, the density is self-determined by
   the pressure $ n= \frac{\partial P}{\partial \mu}|_{T,V} $, so the density and pressure go hand in hand,
   the solid $^{4} He $ is likely to be commensurate.
   We expect that the vacancy-vacancy interaction is also a repulsive dipole-dipole one.
   The condensation of these repulsively interacting vacancies at $ T=0 $ leads to the
   SF mode  inside the  in-commensurate ENS. This resulting state is the ESS state with a lattice constant slightly smaller than $ r_{s} a_{ex} $ to accommodate
   the extra vacancies.  As the distance increases to $ d > d_{c2} > d_{c1}, n_{v}(0)=0 $, the
   resulting state is a commensurate ENS whose lattice constant is locked at $ r_{s} a_{ex} $.
   As distance increases further, the ENS will crossover to the two
   weakly coupled Wigner crystal. It becomes feasible to experimentally explore all
   the possible phases and phase transitions in the EHBL in the near future.

    From the mean field analysis of the Ginsburg-Landau action in \cite{exss}, the lattice structure of the excitonic supersolid
    should be a triangular lattice. When studying the excitations spectrum inside the ESS, a similar GL action can be constructed as its
    3 dimensional counter part \cite{exss}. For a triangular lattice, $ \lambda_{\alpha \beta \gamma \delta}=
    \lambda \delta_{\alpha \beta} \delta_{\gamma \delta}
    + \mu ( \delta_{\alpha \gamma } \delta_{\beta \delta}+ \delta_{\alpha \delta} \delta_{\beta \gamma
    } ) $ where $ \lambda $ and $ \nu $ are Lame coefficients, $
    \rho^{s}_{\alpha,\beta}= \rho^{s} \delta_{\alpha,\beta}, a_{\alpha,\beta}= a
    \delta_{\alpha,\beta} $. Eqn.\ref{is}, the following equations an Fig.1 apply.
    Note that the isotropic 3d solid discussed in Sec.II-1 is just a simplification.
    The supersolidons in 3d hcp crystal were discussed in \cite{epl}. While Eqn.\ref{is} holds rigorously for a 2d triangular lattice.
    The differences between 3d and 2d cases were discussed in \cite{exss}.

\section{ Density wave supersolids and Valence bond Supersolids in lattice systems }

     The extended boson Hubbard model ( EBHM ) with various kinds of interactions,
     on all kinds of lattices and at different filling factors is described by the following
     Hamiltonian:
\begin{eqnarray}
   H_{EBHM}  & = & -t \sum_{ \langle ij \rangle } ( b^{\dagger}_{i} b_{j} + h.c. )
          - \mu \sum_{i} n_{i} + \frac{U}{2} \sum_{i} n_{i} ( n_{i} -1 )
                                  \nonumber   \\
      & + & V_{1} \sum_{ \langle ij\rangle } n_{i} n_{j}  + V_{2} \sum_{ \langle\langle ik\rangle\rangle } n_{i} n_{k} + \cdots
\label{boson}
\end{eqnarray}
    where $ n_{i} = b^{\dagger}_{i} b_{i} $ is the boson density, $ t $ is the nearest neighbor hopping amplitude.
    $ U, V_{1}, V_{2} $ are onsite, nearest neighbor (nn) and next nearest neighbor (nnn) interactions respectively,
    the $ \cdots $ may include further neighbor interactions, dipole-dipole interaction $ V_d= \frac{p^{2}}{| \vec{r}_i -\vec{r}_j |^{3} } $
    and possible ring-exchange interactions.
     A supersolid in Eqn.\ref{boson} is defined as the
     simultaneous orderings of ferromagnet in the $ XY $ component ( namely, $ \langle b_{i} \rangle \neq 0 $ )
     and CDW in the $ Z $ component. The EBHM, especially the stability of the supersolid phase has been studied by
     spin wave expansion \cite{gan}, the Quantum Monte-Carlo (QMC) simulations \cite{square,squaresoft,qmc,sca,sstri,frusqmc} and the dual
     vortex method (DVM) \cite{pq1,mob,bipart,frus1,frus2} which is a Ginsburg-Landau ( GL ) action in the dual vortex picture.

\begin{figure}
\hspace{-0.5cm}
\includegraphics[width=10cm]{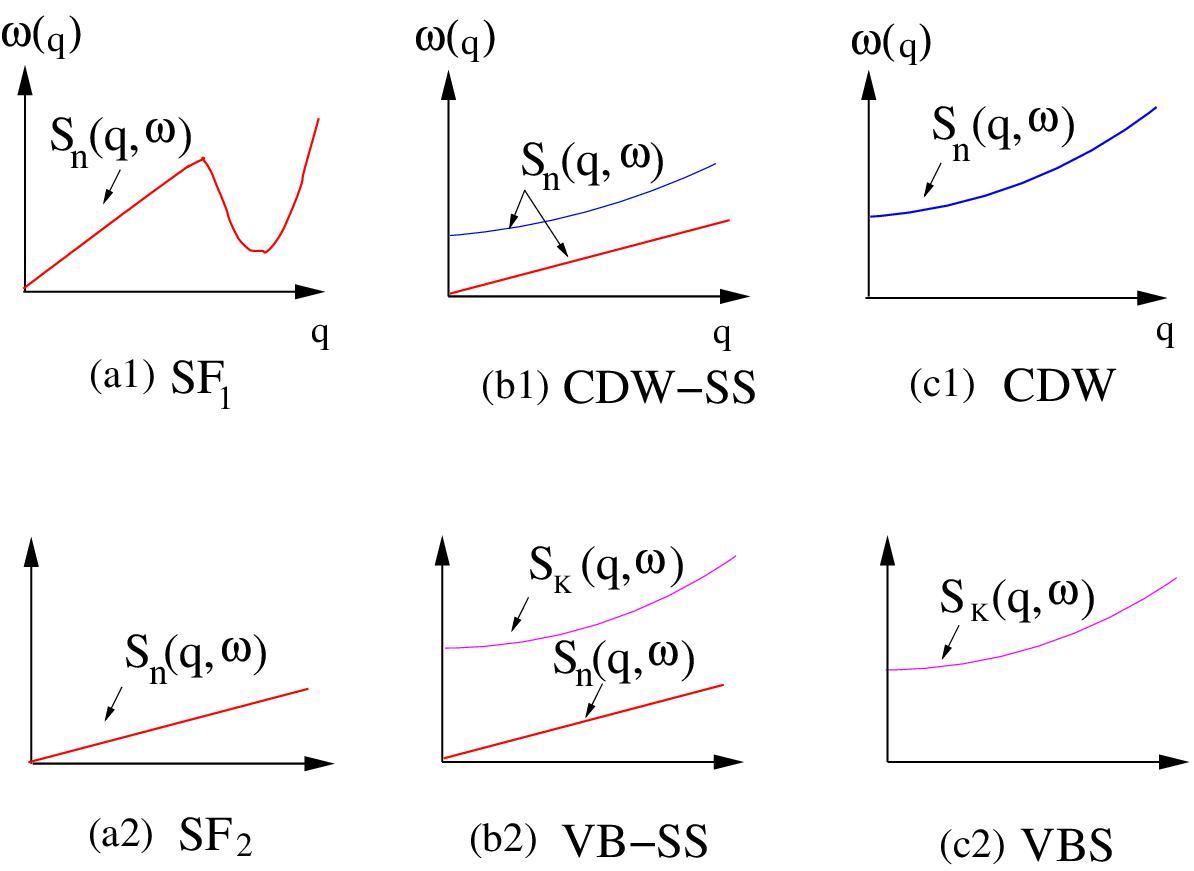}
\caption{ (Color figure online) The excitation spectra in the CDW, VBS, SF, CDW-SS and
VB-SS states in a lattice. They correspond to the peak positions of the corresponding
dynamic response functions shown with arrows \cite{bragg2}. In the (b1) and (c1)
cases, the starting wavevector is $ \vec{Q}_{n} $ in the upper CDW
branch. In the (b2) and (c2) cases, the starting wavevector is $
\vec{Q}_{K} $ in the upper VBS branch.
The corresponding spectral weights are worked out in \cite{bragg2}.}
\label{excitations}
\end{figure}

\subsection{ The Dual Ginsburg-Landau approach: the dual vortex method to study supersolids in lattices }

    In \cite{bipart}, based on the dual vortex degree of freedoms \cite{pq1}, the author developed a systematic dual Ginsburg-Landau (GL) action
    to study all the possible phases and phase transitions in the EBHM Eqn.\ref{boson} in
    bipartite lattices such as a honeycomb and square  lattice near half filling. The dual GL theory can be used to derive the symmetry breaking patterns
    of various insulating and supersolid states, the tarnsitions among different phases and excitation spectrum in a given insulating phase, especially
    in various kinds of supersolids. In the insulating side, it was found that there are two consecutive transitions at zero temperature driven by the chemical potential: in the Ising limit, a  Commensurate-Charge Density Wave (CDW) at half filling to  a narrow window of CDW supersolid (CDW-SS), then to
    an Incommensurate-CDW ; in the easy-plane limit, a
    Commensurate-Valence Bond Solid (VBS) at half filling to
    a narrow window of VBS supersolid ( VB-SS), then to an Incommensurate-VBS.
    The first transition is second order  in the same universality class as the Mott to insulator transition, therefore has the exact critical exponents $
    z=2, \nu=1/2, \eta=0 $ with logarithmic corrections\cite{int}, while the second one is first order.
    The VB-SS is a new kind of solid which only happens on a lattice, so no analog in a continuous system discussed in the Sec.II.
    The VB-SS maybe stabilized in the presence of ring-exchange terms \cite{sandvik}. The excitation spectra in these phases are shown in the Fig.2.
    The transition from the the SF to the CDW-SS transition is driven by the condensation of
    diagonal vortex-antivortex pair without the condensations of the individual vortex \cite{int}. In the direct boson picture, it may correspond to the
    gap closing at the roton minimum.  The transition from the the SF to the VB-SS transition is driven by the condensation of
    off-diagonal vortex-antivortex pair without the condensations of the individual vortex.
    Unfortunately, it is still not clear what kind of physical processes it corresponds to in the direct boson picture \cite{int}.
    Very recently, the author in \cite{frus1,frus2} also studied various kinds of supersolids in
    frustrated lattices such as triangular and
    Kagome lattices . As first discovered in \cite{gan}, a CDW supersolid is very robust in a triangular lattice slightly
    away the $ 1/3 $ filling. It could be either vacancy-like or interstitial-like supersolid. But there is no VB-SS which was discovered in
    bi-partite lattices. There could also be a CDW-VS supersolid which has the three kinds of orders: CDW-order, VB order and the superfluid order.
    In a Kagome lattice, there is no VB-SS either, but there could be CDW-SS and the CDW-VB-SS. So the VB-SS is unique to bi-partite lattices,
    while the CDW-VB-SS is unique to frustrated lattices. While, the CDW-SS can happen in both bi-partite and frustrated lattices.

        Just like the GL approach to the possible supersolids in continuous systems discussed in Sec.I,
        the DVM developed in \cite{pq1,mob,bipart,frus1,frus2} is a  symmetry-based approach
        which, in principle, can be used to classify all the possible phases, especially supersolid phases,
        and phase transitions. But if a particular phase identified by the DVM will become a stable
        ground state or not depends on the specific competitions among all the
        possible parameters in the EBHM in Eqn.\ref{boson}. This  kind of question can only be addressed by a microscopic approach such as Quantum Monte-Carlo
        (QMC) simulations on a specific Hamiltonian. In the following, we just compare with QMC simulations in
        $ V_1, V_2 $ models and the dipole-dipole interaction model $ V_d= \frac{p^{2}}{| \vec{r}_i -\vec{r}_j |^{3} } $.

\subsection{ Quantum Monte-Carlo simulations in lattice models  }

 Due to the lack of sign problems in the EBHM Eqn.\ref{boson}, Numerical calculations such as Quantum Monte-carlo simulations are
 very successful in studying the EBHM with various kinds of long range interactions. The combinations of the DVM discussed in the last subsection
 and the QMC simulations to be discussed in this subsection lead to converging pictures of various supersolids in lattice systems.

{\sl 1.  QMC simulations in lattices with $ V_1, V_2 $ interactions }

  There have been extensive QMC on the EBHM Eqn.\ref{boson}, to especially search for stable supersolid phases  in various
  bipartite and frustrated lattices\cite{square,squaresoft,qmc,sstri,sca,frusqmc}.
  For  hard core bosons in a square lattice, it was shown by the QMC in \cite{square}
  that the $ ( \pi, \pi ) $ X-CDW SS slightly away from $1/2 $ filling is not stable against
  phase separation with  $ U= \infty,  V_1 > 0, V_2 = 0 $, but
  the $ ( \pi, 0 ) $ stripe SS is indeed stable with  $ U= \infty, V_1=0, V_2 > 0 $.
  The transition from the stripe SS to the SF is a first order transition\cite{int}. For soft core bosons with $ U < \infty,  V_1 > 0, V_2 = 0 $,
  the interstitial-like supersolid slightly above  $1/2 $ filling is stable, although the vacancy-like supersolid slightly
  below  $1/2 $ filling is still unstable against phase separation \cite{squaresoft}.
  Similar phenomena were also found for soft core bosons in a honeycomb lattice near half fillings\cite{qmc}.
  The claim that the CDW to the CDW-SS transition  at $ d=2 $ driven by the chemical potential is in the same universality class of SF to
  Mott transition with the critical exponents $ z=2, \nu=1/2, \eta=0 $ reached from the DVM \cite{bipart} was indeed confirmed by the QMC in \cite{sca}.
  Possible supersolids were also studied in a $ d=1 $ lattice \cite{sca}. At $ d=1 $, the SF to the CDW transition
  is in the Kosterlitz-Thouless (KT) transition universality class instead of the first  order transition in $ d=2 $ and $ d=3 $.
  We expect that the SF to the CDW-SS transition
  at $ d=1 $ is also in the KT transition universality class.
  The QMC simulations in a triangular lattice \cite{sstri} found stable supersolids near $ 1/3 $ filling even for hard core
  bosons with $ U= \infty, V_1 > 0 $, as first predicted
  by the spin wave expansion in \cite{gan}.
  However, for hard core bosons with $ U= \infty, V_1 > 0 $, no stable supersolids were found near $ 1/3 $ fillings in a Kagome lattice \cite{frusqmc}.
  But we do expect \cite{frus1,frus2} that a CDW-VB supersolid should be stable in a soft core boson case with $ U < \infty, V_1 > 0 $.
  Furthermore, a stripe supersolid should be stable even in a hard core case with  $ U= \infty, V_1 > 0, V_2 > 0 $.

{\sl 2.  QMC simulations in optical lattices with Dipolar bosons  }

 Since the experimental realization of polar fermionic molecules \cite{junpolar} $ ^{40}K+ ^{87} Rb $, there have been
 extensive research activities to study new states of matter which can be formed by these polar molecules \cite{dipolarss,dipolarsstri,eplss}.
 Stable bosonic molecules $ ^{39} K+ ^{87} Rb $ should also be within experimental reach  in the near future.
 The particular feature of these polar molecules are that they carry large
 electric dipole moments, therefore interact with each other via long-rang anisotropic dipole-dipole interactions similar to excitons in the EHBL in Sec.II-2.
 It was argued in Sect.II-2 that the dipole-dipole interaction between indirect excitons
 may favor a formation of vacancy-like exciton supersolid in some intermediate distances between the bilayers.
 Here, there are extensive numerical evidences that
 the dipole-dipole long-range interaction is especially favorable to the formation of the CDW supersolid \cite{dipolarss,dipolarsstri,eplss}.
 Although the QMC simulations in \cite{square} found that for hard-core bosons in a square lattice with the $ V_1 >0 $ interaction,
 the X-CDW is not stable against a phase separation slightly away from $ 1/2 $ filling, the QMC simulations in \cite{dipolarss} found that
 for hard core bosons with the dipole-dipole interaction, the X-CDW is stable in a large parameter regime
 slightly away from $ 1/2 $ fillings.  Furthermore, it was found the CDW-SS to the SF transition is a second order transition
 in the $ 3d $ Ising universality class\cite{int}, instead of a first order transition in the $ V_1 >0 $ case  \cite{square}.
 Very similarly, the $ \vec{Q}_n= 2 \pi/3 (1,1) $  X-CDW supersolid
 were found to be stable in a large parameter regimes near the $1/3 $ filling in a triangular lattice
 \cite{dipolarsstri}. Stable supersolid phase was also identified in  dipolar bilayer systems \cite{eplss}.
 In fact, the $ ^{52}Cr $ atoms \cite{cromium}
 carry exceptionally large magnetic dipole moment and therefore interact with each other also with
 the anisotropic long-range dipole-dipole interaction.  All kinds of CDW and CDW supersolids could be very likely realized
 in near future experiments with either dipolar bosons or $ ^{52}Cr $ atoms  loaded in square and triangular lattices.

\subsection{ Detection of supersolids and their excitations in optical lattices }

 There could be many kinds of detection methods of possible supersolids in continuous systems such as the NCRI \cite{chan,annealing},
 mass flow \cite{massflow,massflow1}, melting curve \cite{melt}, acoustic attenuation \cite{acou}, specific heat \cite{heat} and X-ray scattering.
 So far, the detection methods of the possible charge neutral cold atoms loaded in optical lattices are very limited. However, in recent works \cite{bragg1,bragg2},  the authors  developed a systematic and unified theory of
   using the optical Bragg scattering, atomic Bragg scattering or cavity QED to detect the ground state and the excitation spectrum
   of many quantum phases of interacting bosons loaded in bipartite and frustrated optical lattices.
   They showed that the two photon Raman transition processes  in the three detection methods not only couple to
   the density order parameter, but also the  {\sl valence bond order } parameter due to the hopping of the bosons
   on the lattice. This valence bond order coupling is very
   sensitive to any superfluid order or any Valence bond (VB ) order
   in the quantum phases to be probed.  These quantum phases include not only the well known superfluid and
   Mott insulating phases, but also other important phases such as various kinds of charge density waves (CDW),
   valence bond solids (VBS), CDW-VBS phases with both CDW and VBS orders unique to frustrated lattices,
   and also various kinds of supersolids. So if the supersolids of dipolar bosons or  $ ^{52}Cr $ atoms can indeed be realized
   in optical lattice, the light scattering methods discussed in \cite{bragg1,bragg2} could be used to detect their
   existence and excitation spectra shown in Fig.2.

\section{ Superfluid density wave (SFDW)  }

  As stressed in the previous two sections, for supersolids in both continuous and lattice systems,
  there are always a underlying normal solid components $ \delta n $ in Eqn.\ref{orderhe}. So in continuous system, there must be vacancies
  in an in-commensurate solid flowing through the whole lattice to form a supersolid. While on lattices, there must be
  vacancies or interstitials away from commensurate fillings to stabilize a lattice supersolid. In this section, we discuss
  Superfluid density wave (SFDW) which has no such  underlying normal solid components $ \delta n $. Its lattice structure is completely
  due to the modulations of the order parameter $ \psi $ in Eqn.\ref{orderhe} itself. So SFDW can happen at any filling factors.

\subsection{ SFDW in continuous systems }

  So far, we only discussed possible supersolids and their excitation spectra in  bosonic systems.
  In fact, there is also an analog in fermionic systems which is the well known,
  but putative Larkin-Ovchinnikov-Fulde-Ferrell (LOFF) pairing  state \cite{ff,lo}.
  When the number of spin up electron is equal to the number of down spin electron $ n_{\uparrow}=n_{\downarrow} $, then the pairing between them
  is at $ \vec{k}=0 $ only. If there is a mismatch $ \delta n= n_{\uparrow} -n_{\downarrow} $, then pairing may shift to a non-zero momentum
  $ q_0 \sim k_{F \uparrow}- k_{F \downarrow} $  which is the FFLO pairing state.   By using the GL theory near the transition from the normal to the FFLO state,
  at a mean field level, the authors in \cite{fflo} constructed the GL free energy in the momentum space in terms of the S-wave pairing order parameter
  $ \psi_{FFLO}(\vec{x} )= \langle c^{\dagger}_{\uparrow} (\vec{x} ) c_{\downarrow}(\vec{x} ) \rangle $:
\begin{eqnarray}
 F_{FFLO} &= & \sum_{G}\frac{1}{2}r_{G}|\psi_{G}|^{2}+u\sum_{G}\psi_{G_{1}}\psi_{G_{2}}\psi_{G_{3}}\psi_{G_{4}}\delta_{G_{1}+G_{2}+G_{3}+G_{4}}\nonumber \\
&+&v\sum_{G}\psi_{G_{1}}\psi_{G_{2}}\psi_{G_{3}}\psi_{G_{4}}\psi_{G_{5}}\psi_{G_{6}}\delta_{G_{1}+
G_{2}+G_{3}+G_{4}+G_{5}+G_{6}}
\label{mom}
\end{eqnarray}
   where $ r_{G}= T-T_{c} $ and $ u,v $ are the fourth and sixth order interaction terms respectively.
   This equation should be understood as an expansion in terms of
   the FFLO order parameter $ \psi_{G} $, not as a gradient expansion anymore.
   The GL action was used to understand the lattice structure of the FFLO state.
   If $ r_{G} > 0 $,  the system is in the normal state with $  \langle \psi(\vec{G}) \rangle  =0 $,
   while when $ r_{G} < 0 $, it is in the FFLO phase with the order parameter:
   $ \langle \psi ( x ) \rangle  = \sum^{P}_{i=1} \Delta_{i}  e^{i \vec{G}_{i} \cdot \vec{x} }, |\vec{G}_{i}|= q_{0} $.
   From Eqn.\ref{mom}, the authors found the most favorable lattice structures of the FFLO state
   is the stripe state (LO state with $ P=2 $ ) in large number of parameter regimes. The FFLO state maybe considered as a weak coupling ( or fermionic ) analog
   of the ( bosonic ) supersolid.

   So far all the previous  analysis  in a FFLO state\cite{ff,lo,fflo} are only at a mean field level.
   Just from symmetry breaking point of views, the FFLO state breaks both $ U(1) $
   symmetry and the translational symmetry, therefore it also supports two kinds of Goldstone
   modes. (1) The Goldstone mode due to the $ U(1) $ symmetry breaking.
   (2) the lattice phonon modes due to the translational symmetry  breaking.
   Above the mean field solution, very similar to the second equation in Eqn.\ref{orderhe}, the pairing order parameter can be written as:
\begin{eqnarray}
 \psi_{FFLO} ( \vec{x}, \tau )  =  \Delta e^{ i \theta(  \vec{x}, \tau ) }  \sum^{\prime}_{\vec{G}}
      e^{i \vec{G} \cdot
     ( \vec{x} + \vec{u}( \vec{x}, \tau ) )}
\label{orderfflo}
\end{eqnarray}
     where $ \theta(  \vec{x}, \tau )  $ and $ \vec{u}( \vec{x}, \tau )  $ are the superfluid phonon and the lattice phonon modes respectively.
     The LO state corresponds to $ P=2 $.
     For a charged condensed matter system such as a electron system, due to the Higgs mechanism,
     the Goldstone mode $ \theta(  \vec{x}, \tau ) $ will be just eaten by the gauge field.
     However, for a neutral system such as  the pairing between two species of fermions with unequal
     populations in  cold atom systems across a Feshbach resonance, the Goldstone mode $ \theta(  \vec{x}, \tau ) $ survives.
 The coupling between the two phonon modes in Eqn.\ref{orderfflo} are also described by a equation similar to Eqn.\ref{is}. The only difference is
 that the lattice structure is a LO state instead of an isotropic solid. After taking this difference into account, the elementary excitations
 inside the FF state  are similar to those in the Fig.1b and the corresponding spectral weights can be worked out similarly. The experimental signatures of the
 elementary excitations can also be worked out similarly.
 Unfortunately, so far, there is still no convincing evidences for a FFLO state in either condensed matter or cold atom systems yet.

\subsection{ SFDW inside an optical cavity }

 As shown in Sec.III, various kinds supersolids may be stabilized in the presence of long-range interactions.
 In this section, we discuss a superfluid density wave (SFDW) of bosons where the long range interactions between bosons
 are meditated by cavity photons. The experimental set-up is shown in Fig.3a where cold atoms such as $ ^{87} Rb $ are embedded in a high
 finess standing wave cavity and are strongly interacting with cavity photons, subject to a transverse pumping.
 There are two kinds of complementary measurements. One is the probe shown in Fig.3a which detects the
 Florescence spectrum of small cavity leaking photons \cite{orbitalthermal,orbital}. Another is the absorption imaging which detects
 the atomic distribution inside the cavity.
 In a frame rotating with the pumping frequency $ \omega_p $ in the Fig.3a,
 the experimental set-up in Fig.3a can be mapped to the $ Z_2 $ Dicke model \cite{berryphase} Eqn.\ref{z2}:
\begin{equation}
  H_{Z_2}  =  \omega_{c} a^{\dagger} a +  \frac{\omega_{a}}{2} \sum^{N}_{i=1} \sigma^{z}_{i}
   + \frac{g}{\sqrt{N}}\sum^{N}_{i=1} ( a^{\dagger} + a ) ( \sigma^{+}_{i}  + \sigma^{-}_{i}  )
\label{z2}
\end{equation}

 In the frame rotating with the pumping frequency $ \omega_p $, the two photon Raman process  in the experimental
 set-up Fig.3a leads to:
\begin{eqnarray}
 H_{RS} & = & \delta a^{\dagger} a + \int dx dz  \Psi^{\dagger}(x,z)[  \frac{ p^{2}_{x} + p^{2}_{z} }{ 2 m } +
 \frac{ \Omega^{2} }{ \Delta_a } \cos^{2} k z    \nonumber  \\
 & + & \frac{ g_0 \Omega }{ \Delta_a } ( a^{\dagger}  + a ) \cos kx  \cos k z
  +  \frac{ g^{2}_0 }{ \Delta_a } \cos^{2} k x a^{\dagger} a ] \Psi (x,z)
\label{z2pump}
\end{eqnarray}
  where $ \delta= \omega_p - \omega_c, \Delta_a= \omega_p- \omega_a $.
  The $ Z_2 $ symmetry is $ a \rightarrow -a, kx \rightarrow kx +
  \pi $ or $ a \rightarrow -a, k z \rightarrow k z +  \pi $.

   At the mean field level, in the two modes approximation, one can decompose the atom field into the superpositions of
   two momentum (orbital) levels:
\begin{equation}
  \Psi_{RS} (x,z) = c_0 \psi_0 + c_1 \cos kx  \cos k z \psi_0
\label{field}
\end{equation}
  where the $ \psi_0 $ is the zero crystal momentum state of the lowest Bloch band of the one dimensional Hamiltonian
  $ H_{1d}= \frac{ p^{2}_{z} }{ 2 m } +
 \frac{ \Omega^{2} }{ \Delta_a } \cos^{2} k z $ and $ c^{\dagger}_0 c_0 + c^{\dagger}_1 c_1= N $. Under the $
  Z_2 $ transformation, $ c_1 \rightarrow - c_1 $. When comparing Eqn.\ref{field} with Eqns.\ref{orderhe},\ref{orderfflo},
  noting that $ \lambda_p = 2 \pi/k, \lambda=\lambda_p/2=\pi/k $ in the Fig.3, one can see that the $ c_1 $ term corresponds to
  $ P=4 $ with the ordering wavevectors $ \vec{G}=(\pm \pi, \pm \pi) $.

\begin{figure}
\includegraphics[width=5cm]{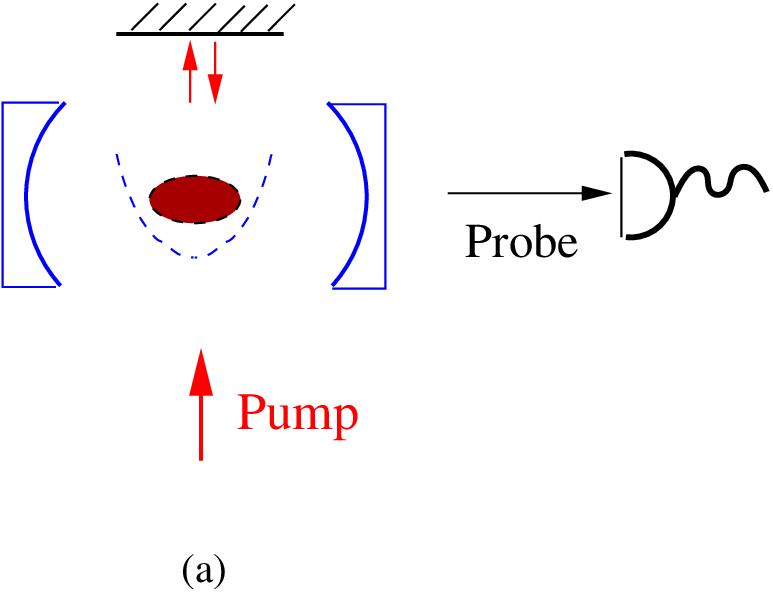}
\hspace{0.5cm}
\includegraphics[width=4cm]{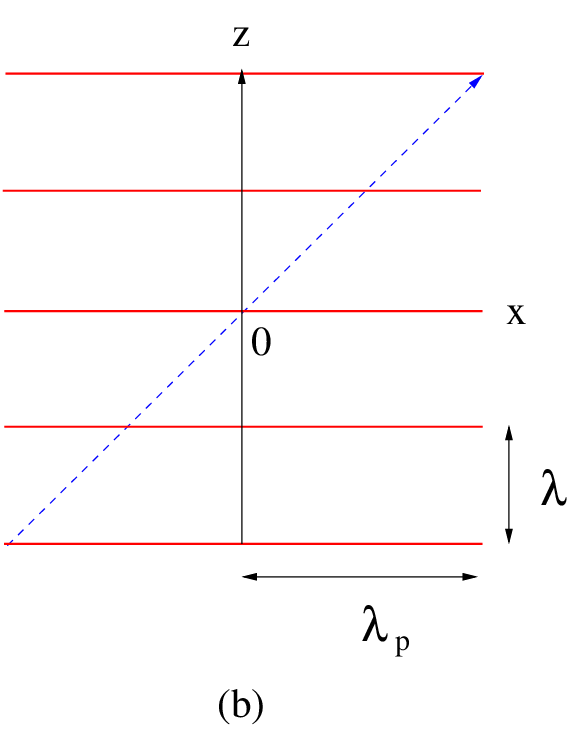}
\hspace{0.5cm}
\includegraphics[width=4cm]{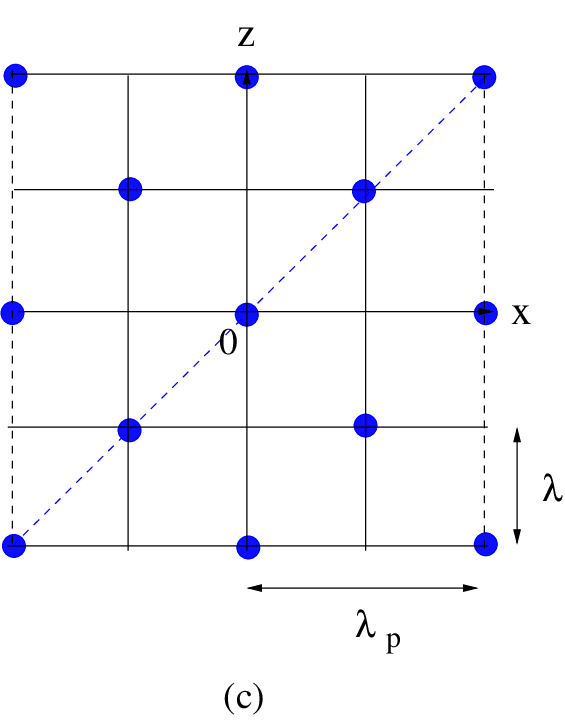}
\caption{ (Color figure online) (a) Reflected Transverse pumping plus the Standing wave
cavity to realize the $ Z_2 $ Dicke model, the probe detects the
Florescence spectrum.  The atom distributions of the $ Z_2 $ Dicke in (b) and (c) can be detected by the absorption imaging. (b) In the normal phase, the atoms just
follow the stripes formed by the reflected transverse pumping. The $ \lambda_p = 2 \pi/k, \lambda=\lambda_p/2=\pi/k $ where the $ k $ is the transverse pumping
wave number (c)
In the $ Z_2 $ superradiant phase for photons and the $ Z_2 $ supersolid state for atoms, the atoms take the check-board
distribution on the optical lattice formed by the cavity photon and the pumping laser. }
\label{z2lattice}
\end{figure}

  Substituting Eqn.\ref{field} into Eqn.\ref{z2pump} leads to the
  interacting Hamiltonian between the photon and the effective two momentum ( orbital ) levels:
\begin{equation}
 H_{J-Z_2 } =\omega_c a^{\dagger} a + \omega_a J^{z} +  \tilde{g} ( a^{\dagger}  + a ) ( J^{-} +  J^{+} )
\label{intz2}
\end{equation}
  where $ \omega_c= \delta- N \frac{ g^{2}_0 }{ 2 \Delta_a }, \omega_a= 2 \omega_r $ where $ \omega_r =\hbar^{2} k^{2}/2 m $ is the recoil energy.
  The effective interaction is $ \tilde{g}= \frac{ g_0 \Omega }{ \Delta_a }  $ and $ J^{z}= \frac{1}{2}
  ( c^{\dagger}_1 c_1-  c^{\dagger}_0 c_0 ), J^{+}=c^{\dagger}_0 c_1, J^{-}=c^{\dagger}_1  c_0 $.
  One can identify Eqn.\ref{intz2} with Eqn.\ref{z2} after
  identifying  the collective spin operators of the $ N $ atoms as $
  J^{z}=  \frac{1}{2} \sum^{N}_{i=1} \sigma^{z}_{i}, J^{+}=\sum^{N}_{i=1}
  \sigma^{+}_{i}, J^{+}=\sum^{N}_{i=1} \sigma^{+}_{i} $.

  From Eqn.\ref{field}, one can see the boson density:
\begin{equation}
  n_{RS}( x, z )  =  \langle c^{\dagger}_0 c_0 \rangle |\psi_0|^{2}+ \langle c^{\dagger}_1 c_1 \rangle \cos^{2} kx \cos^{2} k
  z |\psi_0|^{2}  +  \langle  J^{x}  \rangle \cos k x \cos k z |\psi_0|^{2}
\end{equation}
  In the normal phase,  $ \langle c^{\dagger}_1 c_1 \rangle = 0 $ and $ \langle J^{x} \rangle = 0 $, the boson
  density is just $ \sim |\psi_0|^{2} $ shown in Fig.3b.
  In the  $ Z_2 $ super-radiant phase,  $ \langle c^{\dagger}_1 c_1 \rangle > 0 $ and $ \langle J^{x} \rangle \neq 0
  $, the boson density takes the check-board pattern shown in Fig.3c. The corresponding SF to the SFDW transition of the atoms is
  in the same university class of the $ Z_2 $ superradiance \cite{berryphase}.

  Of course, any symmetry breaking only happens in the thermodynamic limit. For a finite system, the symmetry breaking will be restored.
  The quantum fluctuations ( namely, the finite size effects ) in the $ Z_2 $ Dicke model is
  exponentially suppressed $ \sim e^{- N } $, but still observable
  at small  $ N $. The quantum fluctuations will restore the $ Z_2 $ symmetry, so render  $ \langle  J^{x}  \rangle=0 $,
  but still keep $ \langle c^{\dagger}_1 c_1 \rangle > 0 $ inside the $ Z_2 $ super-radiant phase. They will transform the
  check-board pattern in Fig.3c to the uniform distribution on  the optical lattice formed by the cavity photon and the pumping laser.
  The transition from the SF to the SFDW becomes a crossover.
  The detailed study of quantum fluctuations were given in \cite{berryphase}.

\section{ Conclusions }

    In this paper, we provided a unified and global view on the universal properties of various kinds of supersolids
    such as their trial wavefunctions, symmetry breaking patterns
    and excitation spectra in various systems. These systems could be
    continuous systems such as 3d Helium and 2d excitonic semiconductor systems or various lattice systems such as cold atoms with long range interactions
    loaded on optical lattices.
    Inside a SS phase in a continuous system, the effective action  Eqn.\ref{is}  controls the
    quantum fluctuations above the mean field lattice structure of $  \delta n(\vec{x}) $ and the condensation structure
    of $ \psi(\vec{x}) $. The elementary excitations have two longitudinal modes $  \omega_{\pm}=v_{\pm} q $ called "supersolidons" shown in Fig.1. The transverse modes
    in the SS stay the same as those in the Normal solid. The effects of these two supersolidons on the in-elastic X-ray scattering, neutron scattering, acoustic attenuations and specific heat  were discussed in \cite{qglprl,epl,qgllong}. Detecting these "supersolidons " by these experiments
    could be smoking gun experiments to confirm a supersolid in any continuous systems.
    For the first time, we constructed an effective Hamiltonian whose ground state is a supersolid.
    It can be used to derive the supersolidon spectrum in the whole BZ. We also contrasted the wavefunction and the effective Hamiltonian of a supersolid
    with that of a commensurate solid with local quantum tunneling and exchange processes.
    We argued that the dipole-dipole interaction in the in-direct excitons in 2d EHBL may favor the formation of a supersolid  more than the Van der Waals interactions between $ ^{4} He $ atoms.

    Then we discussed the supersolids in lattice systems which are much simpler than its
    continuous counterparts. Much more established theoretical  results from spin wave analysis, dual Ginsburg-Landau  and QMC
    are established in lattice supersolid cases.
    The valence bond supersolid is a new kind of supersolid unique to lattice systems.
    The elementary excitations in some of these phases were shown in the Fig.2 and can
    be contrasted to its continuous counterparts shown in Fig.1.
    Both interstitial or vacancy like supersolids can happen in lattices slightly away from $ 1/2 $ fillings in a bipartite lattice
    or $ 1/3 $ fillings in a frustrated lattice. So they can only happen at in-commensurate fillings.
    In this regard, the lattice supersolids are similar to its continuous
    counterparts. The CDW-SS could be realized in possible near future experiments with dipolar bosons or  $ ^{52}Cr $ atoms loaded in various optical lattices.

    We also discussed the superfluid density wave (SFDW) in both fermionic and bosonic  systems.
    It has one order parameter $ \psi(x) $ as shown in Eqn.\ref{orderfflo},\ref{field},
    no the underlying normal solid component $ \delta n(\vec{x}) = n(\vec{x})-n_0 $ shown in the first equation in Eqn.\ref{orderhe}.
    This crucial difference than the supersolids lead to the important fact that
    the SFDW can form at any densities instead of just slightly away from commensurate fillings for a supersolid.
    The first example is the inhomogeneous superfluids ( FFLO state ) in a continuous system. Its excitation spectrum is similar to those shown in Fig.1.
    So far, there is no clear experimental evidences of FFLO state in condensed matter or cold atom systems yet, but they are still under
    extensive experimental searches in both
    communities. The second example is the $ Z_2 $ SFDW inside an optical cavity. Its excitation spectrum is similar to those shown in Fig.2.
    There is clear experimental evidences of such a $ Z_2 $ SFDW in cold atoms inside a transversely pumped high finess cavity shown in Fig.3.
    We expect SFDW may also be realized in cold spinor atom BEC systems in the presence of strong spin-orbit coupling generated by
    artificial non-abelian gauge potentials \cite{sorev}. The spin-orbit coupling may lead to spontaneous translational symmetry breaking, the theory
    presented in Sect.IV may apply to such a case. We expect theoretical investigations and experimental searches for
    various kinds of supersolids in various systems are still active underway.

{\bf Acknowledgements }

 Y.C and Q.S.T's research are supported by NSFC-11074004.
 J. Ye thank A.V. Balatsky for his hospitality during J.Ye's visit at LANL.
 J. Ye's research is supported by
 NSF-DMR-1161497, NSFC-11074173, 11174210, Beijing Municipal Commission of Education under grant No.PHR201107121, at KITP is supported in part by the NSF under grant No. PHY-0551164.


\begin{thebibliography}{99}
\bibitem{int} One can understand these results from Fig.2 intuitively. (1) From Right to the middle in Fig.2,
across the transition from the CDW ( VBS ) to the CDW-SS ( VB-SS), the CDW (VBS) mode has a gap across the transition, so can be integrated out,
the SF mode is the only low energy mode, so the
CDW to the CDW-SS transition is in the same universality class of the Mott to SF transition.
(2) From left to the middle in Fig.2, across the transition from the $ SF_1 $ ( $ SF_2 $ ) to the CDW-SS ( VB-SS),
there are SF modes on both sides, so they may couple to possible critical density fluctuation mode near the transition,
so the nature of the transition is much more complicated,  may be resolved by high precisoin QMC on specific microscopic models.
For example, the $ V_1 > 0, V_2 > 0 $ model Eqn.\ref{boson} in Sec.III-B-1 may be in a different universality class than
the dipole-dipole intercation model in Sec.III-B-2.
\end{thebibliography}
\end{document}